\pdfoutput=1
\RequirePackage{ifpdf}
\ifpdf 
\documentclass[pdftex]{sigma}
\else
\documentclass{sigma}
\fi

\numberwithin{equation}{section}
\begin{document}

\newcommand{\arXivNumber}{1402.4606}

\allowdisplaybreaks

\renewcommand{\thefootnote}{$\star$}

\renewcommand{\PaperNumber}{075}

\FirstPageHeading

\ShortArticleName{Energy Spectrum and Phase Transition of Fermi Gas on Noncommutative Space}

\ArticleName{Energy Spectrum and Phase Transition of Superf\/luid Fermi Gas of Atoms on Noncommutative Space\footnote{This
paper is a~contribution to the Special Issue on Deformations of Space-Time and its Symmetries.
The full collection is available at \href{http://www.emis.de/journals/SIGMA/space-time.html}
{http://www.emis.de/journals/SIGMA/space-time.html}}}

\Author{Yan-Gang MIAO~$^{\dag\ddag}$ and Hui WANG~$^\dag$}

\AuthorNameForHeading{Y.-G.~Miao and H.~Wang}

\Address{$^\dag$~School of Physics, Nankai University, Tianjin 300071, China}
\EmailD{\href{mailto:email@address}{miaoyg@nankai.edu.cn}, \href{mailto:email@address}{wanghuiphy@126.com}}

\Address{$^\ddag$~Interdisciplinary Center for Theoretical Study, University of Science and Technology of China,\\
\hphantom{$^\dag$}~Hefei, Anhui 230026, China}

\ArticleDates{Received February 25, 2014, in f\/inal form July 07, 2014; Published online July 10, 2014}

\Abstract{Based on the Bogoliubov non-ideal gas model, we discuss the energy spectrum and phase transition of the
superf\/luid Fermi gas of atoms with a~weak attractive interaction on the canonical noncommutative space.
Because the interaction of a~BCS-type superf\/luid Fermi gas originates from a~pair of Fermionic quasi-particles with
opposite momenta and spins, the Hamiltonian of the Fermi gas on the noncommutative space can be described in terms of
the ordinary creation and annihilation operators related to the commutative space, while the noncommutative ef\/fect
appears only in the coef\/f\/icients of the interacting Hamiltonian.
As a~result, we can rigorously solve the energy spectrum of the Fermi gas on the noncommutative space exactly following
the way adopted on the commutative space without the use of perturbation theory.
In particular, dif\/ferent from the previous results on the noncommutative degenerate electron gas and superconductor
where only the f\/irst order corrections of the ground state energy level and energy gap were derived, we obtain the
nonperturbative energy spectrum for the noncommutative superf\/luid Fermi gas, and f\/ind that each energy level contains
a~corrected factor of cosine function of noncommutative parameters.
In addition, our result shows that the energy gap becomes narrow and the critical temperature of phase transition from
a~superf\/luid state to an ordinary f\/luid state decreases when compared with that in the commutative case.}


\Classification{53D55; 	81T75; 82D55}

\renewcommand{\thefootnote}{\arabic{footnote}} \setcounter{footnote}{0}

\section{Introduction}\label{Section1}

The development of string theory~\cite{Seiberg} and M-theory~\cite{Connes} has shown that the low energy ef\/fective
theory of superstrings can be described by a~noncommutative f\/ield theory.
Precisely speaking, the low energy ef\/fective theory of D-branes in the background of a~second order anti-symmetric
tensor f\/ield is the Yang--Mills theory.
This leads to extensive interests to a~noncommutative spacetime and its geometry~-- the noncommutative
geometry~\cite{ConnesM}.
It is well-known that the essential distinction between quantum theory and classical theory is noncommutativity between
coordinates and momenta, which gives rise to the Heisenberg uncertainty relation.
On a~noncommutative spacetime, moreover, the noncommutativity between dif\/ferent components of coordinates (momenta)
appears besides that between coordinates and momenta.
If only dif\/ferent components of coordinates do not commute with each other, the algebraic relation takes the
form\footnote{For simplicity, we use the convention $\hbar=1$, except for additional explanations.}
\begin{gather*}
[\hat x_{i},\hat x_{j}]=i\theta_{ij},
\qquad
[\hat p_{i},\hat p_{j}]=0,
\qquad
[\hat x_{i},\hat p_{j}]=i\delta_{ij},
\end{gather*}
where $\theta_{ij}$'s ($i,j=1,2,3$) are noncommutative parameters, and correspond to a~constant anti-symmetric tensor
for the canonical (3-dimensional) noncommutative space.

In 1984 Berry found~\cite{Berry} that besides the well-known dynamical phase there exists a~geometric phase now called
Berry phase during the adiabatic evolution of quantum systems.
Some signif\/icant progress in condensed matter physics reveals~\cite{XiaoM} that the Berry phase plays an important role
in the anomalous Hall ef\/fect and other related phenomena.
The relevant research shows~\cite{Zhong,XiaoJ,XiaoJ+} that the Berry phase is equivalent to the appearance of a~magnetic
monopole on the momentum space of crystals, which results in noncommutativity between dif\/ferent components of
coordinates.

The above demonstration means that a~noncommutative spacetime as a~new concept can be utilized to analyze new physical
phenomena from microscopic to mesoscopic systems.
Although a~lot of studies on noncommutative spacetimes mainly focus on quantum f\/ield theory~\cite{Douglas,Douglas+}, it is
interesting to discuss noncommutative ef\/fects of some solvable models in the framework of quantum mechanics, such as the
degenerate electron gas~\cite{Khan} and superconductor~\cite{Basu} on the Moyal plane\footnote{If only one component of
noncommutative parameters, say $\theta_{12}$, is nonzero, and the other two components $\theta_{23}$ and $\theta_{31}$
are zero, such a~(3-dimensional) noncommutative space is called the Moyal plane.}. However, only the f\/irst order
corrections of the ground state energy level and energy gap are obtained~\cite{Basu, Khan} because the Moyal
product~\cite{Moyal} and the corresponding statistical relation~\cite{twist} are very complicated on the noncommutative
plane.
This brings our motivation that whether nonperturbative ef\/fects of space noncommutativity can be calculated for some
specif\/ic physical system or not.
Fortunately, we f\/ind out such a~quantum system~-- the superf\/luid Fermi gas of atoms.
We note that the scenario of noncommutativity in crystals is dif\/ferent from that in the superf\/luid Fermi gas of atoms.
The purpose to mention the former is to provide an example of the appearance of noncommutativity in mesoscopic systems.
In the present paper we just deal with the theoretical problem and predict a~possible hint of the noncommutative ef\/fect
in the future precise measurement of the energy gap and the critical temperature of phase transition of the superf\/luid
Fermi gas of atoms.

Due to the application of the Feshbach resonance technology~\cite{ChinGrimm, Loftus} to the experiments of cold Fermi
atoms, it has been a~hot topic to explore the Fermi gas of atoms with a~repulsive or an attractive interaction.
When one tunes a~scattering length to be positive (negative) using the Feshbach resonance technology, one produces the
Fermi gas with a~repulsive (an attractive) interaction among atoms.
At a~very low temperature the Fermi atoms with a~repulsive interaction can constitute biatomic molecules and thus the
Bose--Einstein condensate can happen, which has been observed experimentally~\cite{Greiner}.
Conversely, if one tunes a~scattering length to be negative, i.e., there exists an attractive interaction in the Fermi
gas of atoms, it is commonly expected that two Fermionic quasi-particles with opposite momenta and spins probably form
a~Cooper pair, so that a~BCS-type superf\/luid state of Fermi gas exists.
Some hints of this type of superf\/luid states have been observed~\cite{Regal,Zwierlein}, which brings our attention to
the superf\/luid Fermi gas of atoms.
In this paper we investigate such a~quantum system on the canonical noncommutative space and predict some interesting
ef\/fects brought by the space noncommutativity, such as the energy gap being narrow and the critical temperature of phase
transition from a~superf\/luid state to an ordinary f\/luid state decreasing, which may provide some evidences for existence
of a~noncommutative space in future experiments.

In this paper the ef\/fects of space noncommutativity on energy spectrum and phase transition are discussed for the dilute
superf\/luid Fermi gas of atoms with a~weak attractive interaction.
At an extremely low temperature, the attractive interaction originates from the interchange of quasi-particles.
Under this condition, two Fermionic quasi-particles with opposite momenta and spins tendentially constitute a~Cooper
pair~\cite{Bardeen, Cooper} because the Cooper pair has a~lower energy level.
Due to this special kind of interaction, we can prove that the product of the creation and annihilation operators of
a~pair of Fermionic quasi-particles with opposite momenta and spins on the canonical noncommutative space equals the
product of the ordinary creation and annihilation operators on the commutative space, which is fascinating as the
creation and annihilation operators on the noncommutative space satisfy a~complicate twisted anti-commutation
relation~\cite{Balachandran}.
As a~result, the Hamiltonian of the superf\/luid Fermi gas on the noncommutative space can be expressed in terms of the
ordinary creation and annihilation operators related to the commutative space, and the noncommutative ef\/fect appears
only in the coef\/f\/icients of the interacting Hamiltonian.
Correspondingly, we can exactly solve the energy spectrum of the superf\/luid Fermi gas on the noncommutative space~by
following the way adopted on the commutative space.
Dif\/ferent from the results obtained previously in the degenerate electron gas~\cite{Khan} and superconductor~\cite{Basu}
on the noncommutative Moyal plane, we calculate the nonperturbative energy spectrum of the superf\/luid Fermi gas.
In addition, when compared with the superf\/luid Fermi gas on the ordinary (commutative) space, each energy level contains
a~corrected factor of cosine function of noncommutative parameters, which makes the energy gap narrow and further
suppresses the critical temperature of phase transition from a~superf\/luid state to an ordinary f\/luid state.
That is, it needs a~lower critical temperature that the phase transition happens for the Fermi gas on the noncommutative space.

The paper is arranged as follows.
In the next section, the superf\/luid Fermi gas on the commutative space is reviewed brief\/ly, which provides a~background
for our further investigation on the canonical noncommutative space.
The energy spectrum and phase transition of the superf\/luid Fermi gas on this noncommutative space are studied in Section~\ref{Section3}.
At last, Section~\ref{Section4} is devoted to a~short conclusion and perspective.

\section{The superf\/luid Fermi gas on the ordinary space}

In this section we retrospect the BCS-type superf\/luid theory on the ordinary space~\cite{Lifshitz}.
There exists no an ideal Fermi gas as an equilibrium system in the natural world, since all kinds of gas will condense
when the temperature is close to the absolute zero.
But a~dilute Fermi gas can exist as a~meta-stable state with an enough long life.
Incidentally, the model describing a~non-ideal Fermi gas is also suitable to an enough dilute Fermi liquid.

Similar to the superconductivity in metal, the BCS-type superf\/luid can be realized at an extremely low temperature if
there exists an attractive interaction in the Fermi gas.
In order to explain the superf\/luid ef\/fect, Bogoliubov introduced~\cite{Bogoliubov, BogoliubovV} the famous Bogoliubov
transformation and thus consistently derived the energy spectrum for such a~``quantum f\/luid" from the microscopic point
of view.

For a~dilute non-ideal gas system consisting of~$N$ spin-$\frac 1 2$ Fermi atoms, only the two-body interaction should
be considered.
Suppose the interaction is spherically symmetric and does not depend on spins, and the system satisf\/ies the following
conditions
\begin{gather*}
\frac{d}{\lambda}\ll1,
\qquad
\frac{d}{l}\ll 1,
\qquad
\lambda\approx l\ll L,
\end{gather*}
where~$d$ is the scattering length,~$\lambda$ is the de Broglie wavelength,~$l$ is the average distance between
particles, and~$L$ denotes the length of edge of the system.

The Hamiltonian of the Fermi gas with a~weak attractive interaction takes the form
\begin{gather*}
H=H_{0}+H_{\rm int} =\sum\limits_{a=1}^{N}\frac{\vec{p}_a^{\,2}}{2m}+\frac{1}{2}\sum\limits_{a\neq
b}U(\vec{x}_{a}-\vec{x}_{b}),
\end{gather*}
where the f\/irst and second terms describe the free and interacting parts, respectively.
Due to the dilution of the Fermi gas, only the two-body interaction should be taken into account, where the interaction
originates~\cite{Lifshitz} from a~pair of Fermionic quasi-particles with opposite momenta and spins.
The above Hamiltonian can be rewritten in terms of creation and annihilation operators
\begin{gather*}
H_{0}  =  \sum\limits_{\vec{p},{\sigma}} \frac{{\vec{p}}\,^2}{2m}c_{\vec{p},{\sigma}}^{\dag}c_{\vec{p},{\sigma}},
\\
H_{\rm int} = \frac{1}{2}\sum\limits_{{\sigma}_{1}^{\prime},{\sigma}_{2}^{\prime},{\sigma}_{1},{\sigma}_{2}} \int
d^{3}x\psi_{{\sigma}_{1}^{\prime}}^{\dag}(\vec{x})\left[\int
d^{3}y\psi_{{\sigma}_{2}^{\prime}}^{\dag}(\vec{y})U(\vec{x}-\vec{y})
\psi_{{\sigma}_{2}}(\vec{y})\right]\psi_{{\sigma}_{1}}(\vec{x})
\\
\phantom{H_{\rm int}}
 = \frac{1}{2}\sum\limits_{{\sigma}_{1}^{\prime},{\sigma}_{2}^{\prime},{\sigma}_{1},{\sigma}_{2}} \int
\frac{d^{3}p_{1}^{\prime}}{(2\pi)^3}
\frac{d^{3}p_{2}^{\prime}}{(2\pi)^3}\frac{d^{3}p_{1}}{(2\pi)^3}\frac{d^{3}p_{2}}{(2\pi)^3}\,
\langle\vec{p}^{\,\prime}_1 {\sigma}_{1}^{\prime},\vec{p}^{\,\prime}_2
{\sigma}_{2}^{\prime}|U|\vec{p}_{2}{\sigma}_{2},\vec{p}_{1}{\sigma}_{1}\rangle
\\
\phantom{H_{\rm int}=}
\times c_{\vec{p}^{\,\prime}_1 ,{\sigma}_{1}^{\prime}}^{\dag}c_{\vec{p}^{\,\prime}_2 ,{\sigma}_{2}^{\prime}}^{\dag}
c_{\vec{p}_{2},{\sigma}_{2}}c_{\vec{p}_{1},{\sigma}_{1}},
\end{gather*}
where $\psi_{{\sigma}}(\vec{x})=\frac{1}{(2\pi)^3}\int d^{3}p\, e^{-i\vec{p}\cdot \vec{x}}\,c_{\vec{p},{\sigma}}$, and
$\langle\vec{p}^{\,\prime}_1 {\sigma}_{1}^{\prime},\vec{p}^{\,\prime}_2
{\sigma}_{2}^{\prime}|U|\vec{p}_{2}{\sigma}_{2},\vec{p}_{1}{\sigma}_{1}\rangle$ is the two-body interaction matrix element.
If the integration is replaced by summation, the interacting Hamiltonian becomes
\begin{gather*}
H_{\rm int}=\frac{1}{2}\sum\limits_{{\sigma}_{1}^{\prime},\sigma_{2}^{\prime},\sigma_{1},\sigma_{2}}\sum\limits_{\vec{p}^{\,\prime}_1 ,
\vec{p}^{\,\prime}_2 ,\vec{p}_{1},\vec{p}_{2}}
\langle\vec{p}^{\,\prime}_1 {\sigma}_{1}^{\prime},
\vec{p}^{\,\prime}_2 {\sigma}_{2}^{\prime}|U|\vec{p}_{2}{\sigma}_{2},\vec{p}_{1}{\sigma}_{1}\rangle\,c_{\vec{p}^{\,\prime}_1 ,
{\sigma}_{1}^{\prime}}^{\dag}c_{\vec{p}^{\,\prime}_2 ,{\sigma}_{2}^{\prime}}^{\dag}c_{\vec{p}_{2},{\sigma}_{2}}c_{\vec{p}_{1},{\sigma}_{1}},
\end{gather*}
where $c_{\vec{p},{\sigma}}^{\dag}$ and $c_{\vec{p},{\sigma}}$ are the creation and annihilation operators of Fermionic
quasi-particles on the commutative space, respectively.
They satisfy the usual anti-commutation relations
\begin{gather}
\big[c_{\vec{p},{\sigma}}, c_{\vec{p}^{\,\prime} ,{\sigma}^{\prime}}\big]_{+}=0,
\qquad
\big[c_{\vec{p},{\sigma}}^{\dag}, c_{\vec{p}^{\,\prime} ,{\sigma}^{\prime}}^{\dag}\big]_{+}=0,
\qquad
\big[c_{\vec{p},{\sigma}},
c_{\vec{p}^{\,\prime} ,{\sigma}^{\prime}}^{\dag}\big]_{+}=\delta_{\vec{p}\vec{p}^{\,\prime} }\,\delta_{{\sigma}{\sigma}^{\prime}},
\label{eq7}
\end{gather}
where $[\cdot,\cdot]_{+}$ denotes an anti-commutator.

In the BCS-type superf\/luid theory~\cite{Lifshitz}, only a~pair of quasi-particles with opposite momenta and spins has
contributions to the interacting Hamiltonian.
Therefore, the terms with ${\sigma}_{1}={\sigma}_{2}$ or ${\sigma}_{1}^{\prime}={\sigma}_{2}^{\prime}$ are vanishing in
$H_{\rm int}$ when spins are summed.
Due to the conservations of momenta and spins that should be ensured in the states before and after interaction, only
the four possible cases remain
\begin{alignat*}{6}
& ({\rm i})
\quad
&& \sigma_{1}=+\frac{1}{2},
\qquad
&& \sigma_{2}=-\frac{1}{2},
\qquad
&& \sigma_{1}^{\prime}=+\frac{1}{2},
\qquad
&& \sigma_{2}^{\prime}=-\frac{1}{2}, &
\\
& ({\rm ii})
\quad
&& \sigma_{1}=+\frac{1}{2},
\qquad
&& \sigma_{2}=-\frac{1}{2},
\qquad
&& \sigma_{1}^{\prime}=-\frac{1}{2},
\qquad
&& \sigma_{2}^{\prime}=+\frac{1}{2},&
\\
& ({\rm iii})
\quad
&& \sigma_{1}=-\frac{1}{2},
\qquad
&& \sigma_{2}=+\frac{1}{2},
\qquad
&& \sigma_{1}^{\prime}=-\frac{1}{2},
\qquad
&& \sigma_{2}^{\prime}=+\frac{1}{2},&
\\
& ({\rm iv})
\quad
&& \sigma_{1}=-\frac{1}{2},
\qquad
&& \sigma_{2}=+\frac{1}{2},
\qquad
&& \sigma_{1}^{\prime}=+\frac{1}{2},
\qquad
&& \sigma_{2}^{\prime}=-\frac{1}{2}.
&
\end{alignat*}
Because the interaction is usually independent of spins, that is, the spin of a~Fermionic quasi-particle remains
unchanged after interaction, cases ({\rm ii}) and ({\rm iv}) do not contribute to $H_{\rm int}$, either.
Furthermore, as case ({\rm i}) has the same contribution as case ({\rm iii}), the Hamiltonian can be simplif\/ied~as
\begin{gather*}
H=\sum\limits_{\vec{p},{\sigma}}\frac{{\vec{p}}\,^2}{2m}c_{\vec{p},{\sigma}}^{\dag}c_{\vec{p},{\sigma}}+\sum\limits_{\vec{p},\vec{p}\,^{\prime}}
\langle\vec{p}\,^{\prime},\frac{1}{2};
-\vec{p}\,^{\prime},-\frac{1}{2}|U|-\vec{p},-\frac{1}{2};\vec{p},\frac{1}{2}\rangle\,c_{\vec{p}\,^{\prime},\frac{1}{2}}^{\dag}
c_{-\vec{p}\,^{\prime},-\frac{1}{2}}^{\dag}c_{-\vec{p},-\frac{1}{2}}c_{\vec{p},\frac{1}{2}}.
\end{gather*}
This is the Hamiltonian of BCS-type superconductor (superf\/luid with spin-$\frac{1}{2}$)
systems on the commutative space.
Note that the spins are explicitly expressed in the interaction matrix element in order to show that the interaction
originates from a~pair of quasi-particles with opposite momenta and spins.

\section{The superf\/luid Fermi gas on noncommutative space}
\label{Section3}

In this section we discuss the superf\/luid Fermi gas on the canonical (3-dimensional) noncommutative space and explore
the ef\/fects brought by space noncommutativity.
Due to the noncommutativity, the Lorentz invariance on the 4-dimensional spacetime (the rotational invariance on the
3-dimensional space in the nonrelativistic case) is lost.
In order to retrieve such an invariance, the twisted Poincar\'e algebra has been
introduced~\cite{twist,Wess}, and then a~physical system on the canonical noncommutative space can possess
a~Lorentz-type invariance, but the common product of functions of coordinates should be replaced by the Moyal
product~\cite{Moyal}
\begin{gather*}
f(x)\star g(x) \equiv
\left. e^{\frac{i}{2}\theta^{\mu\nu}\frac{\partial}{\partial{x}^{\mu}}\frac{\partial}{\partial{y}^{\nu}}}f(x) g(y)\right|_{x=y},
\end{gather*}
where $\mu,\nu=0,1,2,3$ are Lorentzian indices, while we use $i, j=1, 2, 3$ in our case for the canonical
(3-dimensional) noncommutative space.
In consequence, a~quantum system on the noncommutative space maintains the twisted Poincar\'e
symmetry~\cite{twist,Tureanu}.
To our case, the superf\/luid Fermi gas on the noncommutative space has the rotational invariance due to the twisted
Poincar\'e symmetry.

\subsection{Hamiltonian}

The Hamiltonian of superf\/luid Fermi gas can be written as the sum of the free and interacting parts
\begin{gather*}
H_{\theta}=H_{\theta}^{0}+H_{\theta}^{\rm int},
\end{gather*}
where the free and interacting Hamiltonians can be expressed~\cite{Basu, Khan} as follows
\begin{gather*}
H_{\theta}^{0}  =  \sum\limits_{\vec{p},{\sigma}}
\frac{{\vec{p}}\,^2}{2m}a_{\vec{p},{\sigma}}^{\dag}a_{\vec{p},{\sigma}},
\\
H_{\theta}^{\rm int}  =  \frac{1}{2}\sum\limits_{{\sigma}_{1}^{\prime},{\sigma}_{2}^{\prime},{\sigma}_{1},{\sigma}_{2}}
\int d^{3}x\psi_{{\sigma}_{1}^{\prime}}^{\dag}(\vec{x})\star\left[\int
d^{3}y\psi_{{\sigma}_{2}^{\prime}}^{\dag}(\vec{y})\star U(\vec{x}-\vec{y})
\star\psi_{{\sigma}_{2}}(\vec{y})\right]\star\psi_{{\sigma}_{1}}(\vec{x})
\\
\phantom{H_{\theta}^{\rm int}}
 = \frac{1}{2}\sum\limits_{{\sigma}_{1}^{\prime},{\sigma}_{2}^{\prime},{\sigma}_{1},{\sigma}_{2}} \int
\frac{d^{3}p_{1}^{\prime}}{(2\pi)^3}\frac{d^{3}p_{2}^{\prime}}{(2\pi)^3}\frac{d^{3}p_{1}}{(2\pi)^3}\frac{d^{3}p_{2}}{(2\pi)^3}
\langle\vec{p}^{\,\prime}_1 {\sigma}_{1}^{\prime},\vec{p}^{\,\prime}_2
{\sigma}_{2}^{\prime}|U|\vec{p}_{2}{\sigma}_{2},\vec{p}_{1}{\sigma}_{1}\rangle_{\rm NC}
\\
\phantom{H_{\theta}^{\rm int}=}
\times a_{\vec{p}^{\,\prime}_1,{\sigma}_{1}^{\prime}}^{\dag}a_{\vec{p}^{\,\prime}_2,{\sigma}_{2}^{\prime}}^{\dag}
a_{\vec{p}_{2},{\sigma}_{2}}a_{\vec{p}_{1},{\sigma}_{1}}.
\end{gather*}
Here $\psi_{{\sigma}}(\vec{x})=\frac{1}{(2\pi)^3}\int d^{3}p\,e^{-i\vec{p}\cdot \vec{x}}\,a_{\vec{p},{\sigma}}$,
$\langle\vec{p}^{\,\prime}_1 {\sigma}_{1}^{\prime},\vec{p}^{\,\prime}_2 {\sigma}_{2}^{\prime}|U|\vec{p}_{2}{\sigma}_{2},\vec{p}_{1}{\sigma}_{1}\rangle_{\rm NC}$
is the two-body interaction matrix element of Fermionic quasi-particles on the noncommutative space, and
$a_{\vec{p},{\sigma}}^{\dag}$ and $a_{\vec{p},{\sigma}}$ are the corresponding creation and annihilation operators,
respectively.

Note that the anti-commutation relations of the creation and annihilation operators must be twisted
deformed~\cite{Balachandran,Jong,Hazra,Fiore} according to the twisted Poincar\'e algebra
\begin{gather}
a_{\vec{p},{\sigma}}a_{\vec{p}^{\,\prime}},{\sigma}^{\prime}+e^{i\vec{p}\wedge\vec{p}^{\,\prime}}a_{\vec{p}^{\,\prime},{\sigma}^{\prime}}
a_{\vec{p},{\sigma}} = 0,
\label{eq17}
\\
a_{\vec{p},{\sigma}}^{\dag}a_{\vec{p}^{\,\prime},{\sigma}^{\prime}}^{\dag}+e^{i\vec{p}\wedge\vec{p}^{\,\prime}}a_{\vec{p}^{\,\prime},
{\sigma}^{\prime}}^{\dag}
a_{\vec{p},{\sigma}}^{\dag} = 0,
\label{eq18}
\\
a_{\vec{p},{\sigma}}a_{\vec{p}^{\,\prime},{\sigma}^{\prime}}^{\dag}+e^{-i\vec{p}\wedge\vec{p}^{\,\prime}}a_{\vec{p}^{\,\prime},
{\sigma}^{\prime}}^{\dag}
a_{\vec{p},{\sigma}} = \delta_{\vec{p}\,\vec{p}^{\,\prime}}\delta_{{\sigma}{\sigma}^{\prime}},
\label{eq19}
\end{gather}
where $\vec{p}\wedge\vec{p}^{\,\prime}$ is an exterior product whose def\/inition
is $\vec{p}\wedge\vec{p}^{\,\prime}\equiv p_{i}\theta_{ij}p^{\prime}_{j}$, and repeated subscripts denote summation.

The potential depends only on the distance of two quasi-particles, its Fourier transform takes the form
\begin{gather*}
U(\vec{x}-\vec{y})=\frac{1}{{(2\pi)}^{3}}\int d^{3}k \,\tilde{U}(\vec{k})\,e^{-i\vec{k}\cdot(\vec{x}-\vec{y})}.
\end{gather*}
The nonzero matrix element that satisf\/ies the spin conservation is
\begin{gather*}
\langle\vec{p}^{\,\prime}_1 {\sigma}_{1}^{\prime},\vec{p}^{\,\prime}_2 {\sigma}_{2}^{\prime}|U|\vec{p}_{2}{\sigma}_{2},\vec{p}_{1}{\sigma}_{1}\rangle_{\rm NC}
\\
\qquad
 = \int d^{3}x e^{i\vec{p}^{\,\prime}_1 \cdot\vec{x}}\star \left[\int d^{3}y e^{i\vec{p}^{\,\prime}_2 \cdot\vec{y}}\star
\frac{1}{({2\pi})^{3}} \int d^{3}k\big(\tilde{U}(\vec{k})e^{-i\vec{k}\cdot(\vec{x}-\vec{y})}\big)\star
e^{-i\vec{p}_{2}\cdot\vec{y}} \right] \star e^{-i\vec{p}_{1}\cdot\vec{x}},
\end{gather*}
where the Moyal product~\cite{Douglas,Douglas+} for more than two functions is def\/ined~by
\begin{gather*}
f_{1}(x_{1})\star f_{2}(x_{2})\star \cdots \star f_{n}(x_{n}) \equiv
\prod\limits_{a<b}e^{\frac{i}{2}\theta^{ij}\frac{\partial}{\partial x_{a}^{i}} \frac{\partial}{\partial
x_{b}^{j}}}f_{1}(x_{1}) f_{2}(x_{2})\cdots f_{n}(x_{n}).
\end{gather*}

Now we calculate the Moyal product of functions with~$y$ as a~variable~\cite{Khan}
\begin{gather*}
\frac{1}{({2\pi})^{3}}\int d^{3}k\int
d^{3}ye^{i\vec{p}^{\,\prime}_2 \cdot\vec{y}}\star\big(\tilde{U}(\vec{k})e^{i\vec{k}\cdot\vec{y}}\big) \star
e^{-i\vec{p}_{2}\cdot\vec{y}}
\\
\qquad
 = \frac{1}{({2\pi})^{3}}\int d^{3}k\,
e^{\frac{i}{2}(-\vec{p}^{\,\prime}_2 \wedge\vec{k}+\vec{p}^{\,\prime}_2 \wedge\vec{p}_{2}+\vec{k}\wedge\vec{p}_{2})}
\tilde{U}(\vec{k})\int d^{3}y e^{i\vec{p}^{\,\prime}_2 \cdot\vec{y}}e^{i\vec{k}\cdot\vec{y}}e^{-i\vec{p}_{2}\cdot\vec{y}}
\\
\qquad
 = \int d^{3}k\,e^{\frac{i}{2}(-\vec{p}^{\,\prime}_2 \wedge\vec{k}+\vec{p}^{\,\prime}_2 \wedge\vec{p}_{2}+\vec{k}\wedge\vec{p}_{2})}
\tilde{U}(\vec{k}) \delta\big(\vec{k}-(\vec{p}_{2}-\vec{p}^{\,\prime}_2)\big)
\\
\qquad
 = \int d^{3}k\, e^{-\frac{i}{2}\vec{p}^{\,\prime}_2 \wedge\vec{p}_{2}} \tilde{U}(\vec{k})
\delta\big(\vec{k}-(\vec{p}_{2}-\vec{p}^{\,\prime}_2)\big),
\end{gather*}
where $f(k)\star e^{i\vec{k}\cdot\vec{y}}=f(k)e^{i\vec{k}\cdot\vec{y}}$ has been utilized since $f(k)$ does not depend on~$y$.
Similarly, we can get the Moyal product of functions with~$x$ as a~variable
\begin{gather*}
\int d^{3}k\int d^{3}xe^{i\vec{p}^{\,\prime}_1 \cdot\vec{x}}\star \big(\tilde{U}(\vec{k})e^{-i\vec{k}\cdot\vec{x}}\big)
\star e^{-i\vec{p}_{1}\cdot\vec{x}} =({2\pi})^{3}\int d^{3}k\,e^{-\frac{i}{2}\vec{p}^{\,\prime}_1 \wedge\vec{p}_{1}}
\tilde{U}(\vec{k}) \delta\big(\vec{k}-(\vec{p}^{\,\prime}_1-\vec{p}_{1})\big).
\end{gather*}
As a~result, the interaction matrix element becomes
\begin{gather*}
\langle\vec{p}^{\,\prime}_1 {\sigma}_{1}^{\prime},\vec{p}^{\,\prime}_2 {\sigma}_{2}^{\prime}|U|\vec{p}_{2}{\sigma}_{2},\vec{p}_{1}{\sigma}_{1}\rangle_{\rm NC}
\\
\qquad
 = ({2\pi})^{3} \int d^{3}k\,
e^{-\frac{i}{2}(\vec{p}^{\,\prime}_1 \wedge\vec{p}_{1}+\vec{p}^{\,\prime}_2 \wedge\vec{p}_{2})}\tilde{U}(\vec{k})
\delta\big(\vec{k}-(\vec{p}^{\,\prime}_1 -\vec{p}_{1})\big) \delta\big(\vec{k}-(\vec{p}_{2}-\vec{p}^{\,\prime}_2)\big)
\\
\qquad
 = e^{-\frac{i}{2}(\vec{p}^{\,\prime}_1 \wedge\vec{p}_{1}+\vec{p}^{\,\prime}_2 \wedge\vec{p}_{2})}({2\pi})^{3}\tilde{U}(\vec{p}^{\,\prime}_1-\vec{p}_{1})
\delta\big((\vec{p}^{\,\prime}_1-\vec{p}_{1})-(\vec{p}_{2}-\vec{p}^{\,\prime}_2)\big),
\end{gather*}
where~$\delta$ function presents the momentum conservation.
When the integration is replaced by summation, the interacting Hamiltonian reads
\begin{gather}
H_{\theta}^{\rm int}
 = \frac{1}{2}\sum\limits_{{\sigma}_{1}^{\prime},{\sigma}_{2}^{\prime},{\sigma}_{1},{\sigma}_{2}}
 \sum\limits_{\vec{p}^{\,\prime}_1 ,\vec{p}^{\,\prime}_2,\vec{p}_{1},
\vec{p}_{2}}e^{-\frac{i}{2}(\vec{p}^{\,\prime}_1 \wedge\vec{p}_{1}+\vec{p}^{\,\prime}_2 \wedge\vec{p}_{2})}
\tilde{U}(\vec{p}^{\,\prime}_1-\vec{p}_{1})\delta_{(\vec{p}^{\,\prime}_1-\vec{p}_{1}),(\vec{p}_{2}-\vec{p}^{\,\prime}_2)}
\nonumber
\\
\phantom{H_{\theta}^{\rm int}=}
\times
a_{\vec{p}^{\,\prime}_1 ,{\sigma}_{1}^{\prime}}^{\dag}a_{\vec{p}^{\,\prime}_2 ,{\sigma}_{2}^{\prime}}^{\dag}a_{\vec{p}_{2},{\sigma}_{2}}
a_{\vec{p}_{1},{\sigma}_{1}},
\label{eq26}
\end{gather}
where $({2\pi})^{3}\delta((\vec{p}^{\,\prime}_1-\vec{p}_{1})-(\vec{p}_{2}-\vec{p}^{\,\prime}_2))$ has been
replaced by $\delta_{(\vec{p}^{\,\prime}_1-\vec{p}_{1}),(\vec{p}_{2}-\vec{p}^{\,\prime}_2)}$.

Similar to the case on the commutative space, the interactions in superconductor and superf\/luid mainly originate from
the pairs of Fermionic quasi-particles with opposite momenta and spins, then the Hamiltonian of superf\/luid Fermi gas on
the noncommutative space can be simplif\/ied as
\begin{gather}
H_{\theta}=\sum\limits_{\vec{p},{\sigma}}
\frac{{\vec{p}}\,^2}{2m}a_{\vec{p},{\sigma}}^{\dag}a_{\vec{p},{\sigma}}+\sum\limits_{\vec{p},\vec{p}^{\,\prime} }
\tilde{U}{(\vec{p}^{\,\prime}-\vec{p})}
e^{-i\vec{p}^{\,\prime}\wedge\vec{p}}a_{\vec{p}^{\,\prime},\frac{1}{2}}^{\dag}a_{-\vec{p}^{\,\prime},-\frac{1}{2}}^{\dag}a_{-\vec{p},-\frac{1}{2}}
a_{\vec{p},\frac{1}{2}}.
\label{eq27}
\end{gather}
This Hamiltonian reduces to the form of the commutative case when the noncommutative parameters tend to zero, which
shows the consistency of our noncommutative generalization.
Due to the complications of the twisted anti-commutation relations (see equations~\eqref{eq17},~\eqref{eq18}
and~\eqref{eq19}) and of the multi-body wave functions in the Fock representation~\cite{Jong}, it is quite dif\/f\/icult to
exactly solve the energy spectrum directly from equation~\eqref{eq27}, only some perturbative results can be obtained,
such as the degenerate electron gas~\cite{Khan} and the superconductor~\cite{Basu} on the Moyal plane.

In order to get the nonperturbative energy spectrum for our system (see equation~\eqref{eq27}), we transform
equation~\eqref{eq27} to such a~formulation that is expressed in terms of the creation and annihilation operators on the
ordinary space, i.e., in terms of $c_{\vec{p},{\sigma}}^{\dag}$ and $c_{\vec{p},{\sigma}}$.
Concretely, we use the transformation~\cite{Balachandran, Basu} of the creation and annihilation operators between the
noncommutative and commutative spaces\footnote{The twisted anti-commutation relations
(equations~\eqref{eq17},~\eqref{eq18} and~\eqref{eq19}) can be verif\/ied to be ensured in terms of equations~\eqref{eq7}
and~\eqref{eq28}.}
\begin{gather}
a_{\vec{p},{\sigma}}= c_{\vec{p},{\sigma}}e^{-\frac{i}{2}\vec{p}\wedge\vec{P}},
\qquad
a_{\vec{p},{\sigma}}^{\dag}= e^{\frac{i}{2}\vec{p}\wedge\vec{P}}c_{\vec{p},{\sigma}}^{\dag},
\label{eq28}
\end{gather}
where $\vec{P}$ is def\/ined as the total momentum of the Fermi gas\footnote{It is easy to verify equation~\eqref{eq29}
by using equation~\eqref{eq28}, i.e., the transformation between~$a_{\vec{p},{\sigma}}^{\dag}$,~$a_{\vec{p},{\sigma}}$ and~$c_{\vec{p},{\sigma}}^{\dag}$,~$c_{\vec{p},{\sigma}}$.
This shows that the total momentum operator is same no matter it is expressed by the creation and annihilation operators
on the noncommutative space or by that on the commutative space.}
\begin{gather}
P_{i}\equiv
\sum\limits_{\vec{p},{\sigma}}p_{i}a_{\vec{p},{\sigma}}^{\dag}a_{\vec{p},{\sigma}}=\sum\limits_{\vec{p},{\sigma}}p_{i}
c_{\vec{p},{\sigma}}^{\dag} c_{\vec{p},{\sigma}},
\label{eq29}
\end{gather}
and then express the Hamiltonian (equation~\eqref{eq27}) with the creation and annihilation operators on the commutative
space.
This makes it possible for us to deal with the superf\/luid Fermi gas on the noncommutative space by following the
method~\cite{Lifshitz} adopted on the commutative space.

Using equations~\eqref{eq7} and~\eqref{eq29}, we can easily obtain
\begin{gather*}
c_{\vec{q},{\sigma}}\vec{P}=\vec{P}c_{\vec{q},{\sigma}} +\vec{q}c_{\vec{q},{\sigma}},
\qquad
c_{\vec{q},{\sigma}}^{\dag}\vec{P}= \vec{P}c_{\vec{q},{\sigma}}^{\dag}-\vec{q}c_{\vec{q},{\sigma}}^{\dag}.
\end{gather*}
Again using equation~\eqref{eq28}, we have
\begin{gather*}
a_{\vec{p},{\sigma}}^{\dag}a_{\vec{q},{\sigma^{\prime}}} = e^{\frac{i}{2}\vec{p}\wedge\vec{P}}c_{\vec{p},{\sigma}}^{\dag}
c_{\vec{q},{\sigma^{\prime}}}e^{-\frac{i}{2}\vec{q}\wedge\vec{P}}
 = e^{\frac{i}{2}\vec{p}\wedge\vec{P}}e^{-\frac{i}{2}\vec{q}\wedge(\vec{P}-\vec{p})}c_{\vec{p},{\sigma}}^{\dag}c_{\vec{q},{\sigma^{\prime}}}
 = e^{\frac{i}{2}\vec{q}\wedge\vec{p}}e^{\frac{i}{2}(\vec{p}-\vec{q})\wedge\vec{P}}c_{\vec{p},{\sigma}}^{\dag}c_{\vec{q},{\sigma^{\prime}}}.
\end{gather*}
If $\vec{p}=\vec{q}$, then the above equation is simplif\/ied to be
\begin{gather}
a_{\vec{p},{\sigma}}^{\dag}a_{\vec{p},{\sigma^{\prime}}}=c_{\vec{p},{\sigma}}^{\dag}c_{\vec{p},{\sigma^{\prime}}},
\label{eq32}
\end{gather}
which will be substituted into the free part of equation~\eqref{eq27}.
Similarly, we can derive the following useful formulae
\begin{gather}
a_{\vec{p},{\sigma}}a_{\vec{q},{\sigma^{\prime}}} = c_{\vec{p},{\sigma}}e^{-\frac{i}{2}\vec{p}\wedge\vec{P}}c_{\vec{q},{\sigma^{\prime}}}
e^{-\frac{i}{2}\vec{q}\wedge\vec{P}}
 = c_{\vec{p},{\sigma}}c_{\vec{q},{\sigma^{\prime}}}e^{-\frac{i}{2}\vec{p}\wedge(\vec{P}-\vec{q})}e^{-\frac{i}{2}\vec{q}\wedge\vec{P}}
\nonumber
\\
\phantom{a_{\vec{p},{\sigma}}a_{\vec{q},{\sigma^{\prime}}}}
 = c_{\vec{p},{\sigma}}c_{\vec{q},{\sigma^{\prime}}}e^{\frac{i}{2}\vec{p}\wedge\vec{q}}e^{-\frac{i}{2}(\vec{p}+\vec{q})\wedge\vec{P}},
\label{eq33}
\\
a_{\vec{p},{\sigma}}^{\dag}a_{\vec{q},{\sigma^{\prime}}}^{\dag} = e^{\frac{i}{2}\vec{p}\wedge\vec{P}}c_{\vec{p},{\sigma}}^{\dag}
e^{\frac{i}{2}\vec{q}\wedge\vec{P}}c_{\vec{q},{\sigma}^{\prime}}^{\dag}
 = e^{\frac{i}{2}\vec{p}\wedge\vec{P}}e^{\frac{i}{2}\vec{q}\wedge(\vec{P}-\vec{p})}
c_{\vec{p},{\sigma}}^{\dag}c_{\vec{q},{\sigma}^{\prime}}^{\dag}
\nonumber
\\
\phantom{a_{\vec{p},{\sigma}}^{\dag}a_{\vec{q},{\sigma^{\prime}}}^{\dag}}
 = e^{\frac{i}{2}\vec{p}\wedge\vec{q}}e^{\frac{i}{2}(\vec{p}+\vec{q})\wedge\vec{P}}c_{\vec{p},{\sigma}}^{\dag}c_{\vec{q},{\sigma}^{\prime}}^{\dag}.
\label{eq34}
\end{gather}
Note that a~pair of interacting Fermionic quasi-particles have opposite momenta in the superf\/luid state.
In consequence, we set $\vec{p}=-\vec{q}$ in equations~\eqref{eq33} and~\eqref{eq34} and simplify them to be
\begin{gather}
a_{\vec{p},{\sigma}}a_{-\vec{p},{\sigma^{\prime}}}=c_{\vec{p},{\sigma}}c_{-\vec{p},{\sigma^{\prime}}},
\qquad
a_{\vec{p},{\sigma}}^{\dag}a_{-\vec{p},{\sigma^{\prime}}}^{\dag}=c_{\vec{p},{\sigma}}^{\dag}c_{-\vec{p},{\sigma^{\prime}}}^{\dag},
\label{eq35}
\end{gather}
which will be substituted into the interacting part of equation~\eqref{eq27}.
Equation~\eqref{eq35} means that the product of annihilation (creation) operators of two Fermionic quasi-particles with
opposite momenta on the noncommutative space is same as that on the commutative space, which emerges from the special
interaction in the Fermi gas at low temperature.
Now substituting equations~\eqref{eq32} and~\eqref{eq35} into equation~\eqref{eq27}, we f\/inally obtain
\begin{gather}
H_{\theta}=\sum\limits_{\vec{p},{\sigma}}
\frac{{\vec{p}}\,^2}{2m}c_{\vec{p},{\sigma}}^{\dag}c_{\vec{p},{\sigma}}+\sum\limits_{\vec{p},\vec{p}^{\,\prime}}
\tilde{U}(\vec{p}^{\,\prime}-\vec{p})e^{-i\vec{p}^{\,\prime} \wedge\vec{p}}c_{\vec{p}^{\,\prime},\frac{1}{2}}^{\dag}
c_{-\vec{p}^{\,\prime},-\frac{1}{2}}^{\dag} c_{-\vec{p},-\frac{1}{2}}c_{\vec{p},\frac{1}{2}}.
\label{eq36}
\end{gather}

Equation~\eqref{eq36} is the key point in the present paper.
The Hamiltonian of superf\/luid Fermi gas on the noncommutative space has now been expressed in terms of the creation and
annihilation operators on the commutative space, while the ef\/fect of space noncommutativity appears only in the
coef\/f\/icients of the interacting Hamiltonian.
As a~result, we can calculate the energy spectrum of superf\/luid Fermi gas by following the method used on the
commutative space.
Moreover, if $\theta_{ij}\rightarrow0$, the Hamiltonian equation~\eqref{eq36} reduces to the commutative case, which
shows the consistency of our noncommutative generalization.

We note that the interacting Hamiltonian equation~\eqref{eq26} is the same as equations~(3.8), (3.9) of~\cite{Basu}.
However, both papers do not deal with the same situation.
The reason is that equation~\eqref{eq26} is only a~simple generalization of the interacting Hamiltonian from the
commutative space to the canonical noncommutative space through replacing the common ``product'' by the ``Moyal star
product''.
This treatment is usually adopted for a~system with a~general interaction when one generalizes it from a~commutative
space to a~noncommutative space.
In the present paper the superf\/luid Fermi gas of atoms with a~weak attractive interaction on the canonical
(3-dimensional) noncommutative space can be dealt with nonperturbatively, while in~\cite{Basu} the superconductor
on the Moyal plane can only be treated perturbatively.

In Section~\ref{Section1} of this paper we state our motivation that whether nonperturbative ef\/fects of space noncommutativity exist or not.
To this end, we consider the superf\/luid Fermi gas of atoms with pairs of quasi-particles of opposite momenta and spins,
and indeed f\/ind the nonperturbative Hamiltonian on the noncommutative space as described by equation~\eqref{eq36}.
To our knowledge, if one is limited to search for nonperturbative ef\/fects of space noncommutativity in the superf\/luid
Fermi gas of atoms with a~weak attractive interaction on the canonical noncommutative space, equation~\eqref{eq36} is
a~complete solution for the nonperturbative energy spectrum (see the next subsection).
Nonetheless, if one tries perturbatively, more solutions are possible.

In the above discussion of this subsection, an undeclared but underlying theoretical precondition for the validity of
the work is that Bose and Fermi statistics can be fulf\/illed on the noncommutative spacetime, in spite of the deformed
commutation relations among the coordinates and/or the creation (annihilation) operators.
For the twist-induced noncommutativity this has been f\/irst pointed out in~\cite{Fiore1998, FioreSchupp}, and
applied in~\cite{Fiore2010, Fiore} to twist-induced noncommutative quantum f\/ield theory, in particular on Moyal spaces.
It has been stressed in~\cite{Fiore2010} that the realization (equation~\eqref{eq28}) of the deformed creation
(annihilation) operators in terms of undeformed ones allows $*$-representations of the twisted algebra
(equations~\eqref{eq17}--\eqref{eq19}) only on the ordinary Fermi Fock space.
By the way, equation~\eqref{eq28} is the application to the Moyal case~\cite{Fiore2010} of general
formulas~\cite{Fiore1998,Fiore2010} valid for generic twist-induced deformations.
The fact that twisted/deformed (anti)commutation relations do not give rise to any deformation in Fermi--Dirac or
Bose--Einstein statistics has also been demonstrated very clearly in~\cite{BVaidya,YCDevi}.

\subsection{Energy spectrum}

Now we discuss the energy level of the ground state for the superf\/luid Fermi gas on the canonical noncommutative space.
Two Fermionic quasi-particles with opposite momenta and spins form a~Cooper pair at an extremely low temperature.
Similar to the commutative case~\cite{Lifshitz}, the BCS wave function of the ground state on the noncommutative space
takes the form,
\begin{gather}
|{\rm BCS}\rangle
=\prod\limits_{\vec{p}}\Big(u_{\vec{p}}+v_{\vec{p}}\,a_{\vec{p},\frac{1}{2}}^{\dag}a_{-\vec{p},-\frac{1}{2}}^{\dag}\Big)|{\rm Vac}\rangle
=\prod\limits_{\vec{p}}\Big(u_{\vec{p}}+v_{\vec{p}}\,c_{\vec{p},\frac{1}{2}}^{\dag}c_{-\vec{p},-\frac{1}{2}}^{\dag}\Big)|{\rm Vac}\rangle,
\label{eq37}
\end{gather}
where equation~\eqref{eq35} has been used in order to express $|{\rm BCS}\rangle$ in terms of the creation operators of
the commutative space, $u_{\vec{p}}^{2}$ denotes the probability that the Cooper pair is not taken up, while~$v_{\vec{p}}^{2}$ has the opposite meaning, and they satisfy the normalization condition
$u_{\vec{p}}^{2}+v_{\vec{p}}^{2}=1$.
Here $|{\rm BCS}\rangle$ is the ground state wave function of superf\/luid Fermi gas, and $|{\rm Vac}\rangle$ is the
vacuum state and satisf\/ies $c_{-\vec{p},-\frac{1}{2}}|{\rm Vac}\rangle=0$, $c_{\vec{p},\frac{1}{2}}|{\rm Vac}\rangle=0$,
and $c_{\vec{p},\frac{1}{2}}c_{\vec{p},\frac{1}{2}}^{\dag}|{\rm Vac}\rangle=|{\rm Vac}\rangle$.

Using equations~\eqref{eq36} and~\eqref{eq37}, we can get the ground state energy level of superf\/luid Fermi gas on the
noncommutative space,
\begin{gather}
E_{\theta}^0 = \langle{\rm BCS}| H_{\theta}^{\prime}|{\rm BCS}\rangle
 = 2\sum\limits_{\vec{p}}\eta_{\vec{p}}\, v_{\vec{p}}^{2}+\sum\limits_{\vec{p},\vec{p}^{\,\prime}}\tilde{U}(\vec{p}^{\,\prime}-\vec{p})
e^{-i\vec{p}^{\,\prime}\wedge\vec{p}}u_{\vec{p}}\, v_{\vec{p}}\, u_{\vec{p}^{\,\prime}} v_{\vec{p}^{\,\prime}},
\label{eq38}
\end{gather}
where $H_{\theta}^{\prime} \equiv H_{\theta}-\mu_{\rm F} N$, $\mu_{\rm F}=\frac{{{{\vec p}_{\rm F}}}^{\,2}}{2m}$ is the Fermi energy of
a~single quasi-particle and ${{\vec p}_{\rm F}}$ is the Fermi momentum, and~$N$ is the total particle number operator.
Moreover, the def\/inition of~$\eta_{\vec{p}}$~is
\begin{gather}
\eta_{\vec{p}}\equiv \frac{\vec{p}\,^{2}}{2m}-\mu_{\rm F},
\label{eq39}
\end{gather}
which describes the energy of a~single quasi-particle above the Fermi surface.
Considering the isotropy of superf\/luid Fermi gas, $\tilde{U}(\vec{p}^{\,\prime}-\vec{p})$ depends only on the absolute
value of the momentum dif\/ference.
In addition, using the property $u_{\vec{p}}=u_{-\vec{p}}$ and $v_{\vec{p}}=v_{-\vec{p}}$, that is, $u_{\vec{p}}$ and
$v_{\vec{p}}$ also depend on the absolute value of momenta, we can see that the imaginary part of exponential function
in equation~\eqref{eq38} is vanishing.
This ensures that the ground state energy level is real.
In the following discussion we shall see that the energy levels of excited states are real, too.
Thus we get the nonperturbative energy level of the ground state for the superf\/luid Fermi gas on the noncommutative
space
\begin{gather}
E_{\theta}^0
=2\sum\limits_{\vec{p}}\eta_{\vec{p}}\, v_{\vec{p}}^{2}+\sum\limits_{\vec{p},\vec{p}^{\,\prime}}\tilde{U}(\vec{p}^{\,\prime}-\vec{p})
\cos{\left(p_{i}\theta_{ij}p_{j}^{\prime}\right)}u_{\vec{p}} \, v_{\vec{p}} \,u_{\vec{p}^{\,\prime}}\,  v_{\vec{p}^{\,\prime}}.
\label{eq40}
\end{gather}
When compared with the result on the commutative space, $E_{\theta}^0$ contains an extra corrected factor of cosine
function.
Furthermore, it reduces to the result on the commutative space when the noncommutative parameters tend to zero.

Next we calculate the nonperturbative energy levels of excited states for the superf\/luid Fermi gas on the noncommutative
space.
To this end, we use the Bogoliubov transformation~\cite{Lifshitz}
\begin{gather}
b_{\vec{p},\frac{1}{2}} = u_{\vec{p}}\,c_{\vec{p},\frac{1}{2}}+v_{\vec{p}}\,c_{-\vec{p},-\frac{1}{2}}^{\dag},
\label{eq41}
\\
b_{\vec{p},-\frac{1}{2}} = u_{\vec{p}}\,c_{{\vec{p}},-\frac{1}{2}}-v_{\vec{p}}\,c_{-\vec{p},\frac{1}{2}}^{\dag},
\label{eq42}
\end{gather}
where $b_{\vec{p},\sigma}$ ($b_{\vec{p},\sigma}^{\dag}$) is a~new annihilation (creation) operator of quasi-particles.
Using equations~\eqref{eq41} and~\eqref{eq42} and considering equation~\eqref{eq7} and the condition
$u_{\vec{p}}^{2}+v_{\vec{p}}^{2}=1$, we can verify that the anti-commutation relations of the new creation and
annihilation operators coincide with the usual anti-commutation relations as follows
\begin{gather*}
\big[b_{\vec{p},\sigma}, b_{\vec{p}^{\,\prime},{\sigma^{\prime}}}\big]_{+}=0,
\qquad
\big[b_{\vec{p},\sigma}^{\dag}, b_{\vec{p}^{\,\prime},{\sigma^{\prime}}}^{\dag}\big]_{+}=0,
\qquad
\big[b_{\vec{p},\sigma},b_{\vec{p}^{\,\prime},{\sigma^{\prime}}}^{\dag}\big]_{+}=\delta_{\vec{p}\vec{p}^{\,\prime}}\delta_{{\sigma}{\sigma^{\prime}}}.
\end{gather*}
Solving the inverse transformation from equations~\eqref{eq41} and~\eqref{eq42},
\begin{gather*}
c_{\vec{p},\frac{1}{2}} = u_{\vec{p}}\,b_{\vec{p},\frac{1}{2}}+v_{\vec{p}}\,b_{-\vec{p},-\frac{1}{2}}^{\dag},
\qquad
c_{\vec{p},-\frac{1}{2}} = u_{\vec{p}}\,b_{\vec{p},-\frac{1}{2}}-v_{\vec{p}}\,b_{-\vec{p},\frac{1}{2}}^{\dag},
\end{gather*}
and substituting it into equation~\eqref{eq36}, and further considering the symmetry of $\vec{p}$ and $-\vec{p}$ in the
summation, we can obtain
\begin{gather}
H_{\theta}^{\prime}  \equiv  H_{\theta}-{\mu}_{\rm F} N
 = 2\sum\limits_{\vec{p}}\eta_{\vec{p}}\,v_{\vec{p}}^{2}+\sum\limits_{\vec{p}}\eta_{\vec{p}}\big(u_{\vec{p}}^{2}-v_{\vec{p}}^{2}\big)
\Big(b_{\vec{p},\frac{1}{2}}^{\dag}b_{\vec{p},\frac{1}{2}}
+b_{\vec{p},-\frac{1}{2}}^{\dag}b_{\vec{p},-\frac{1}{2}}\Big)
\nonumber
\\
\phantom{H_{\theta}^{\prime}=}
{} + 2\sum\limits_{\vec{p}}\eta_{\vec{p}}\,u_{\vec{p}}\,v_{\vec{p}}\Big(b_{\vec{p},\frac{1}{2}}^{\dag}b_{-\vec{p},-\frac{1}{2}}^{\dag}
+b_{-\vec{p},-\frac{1}{2}}b_{\vec{p},\frac{1}{2}}\Big)
+\sum\limits_{\vec{p},\vec{p}^{\,\prime} }\tilde{U}(\vec{p}^{\,\prime} -\vec{p})e^{-i\vec{p}^{\,\prime} \wedge\vec{p}}B_{\vec{p}^{\,\prime} }^{\dag}B_{\vec{p}},
\label{eq46}
\end{gather}
where $B_{\vec{p}}$ is def\/ined as
\begin{gather}
B_{\vec{p}}\equiv u_{\vec{p}}^{2}\,b_{-\vec{p},-\frac{1}{2}}b_{\vec{p},\frac{1}{2}}
-v_{\vec{p}}^{2}\,b_{\vec{p},\frac{1}{2}}^{\dag}b_{-\vec{p},-\frac{1}{2}}^{\dag}+
u_{\vec{p}}\,v_{\vec{p}}\Big(b_{-\vec{p},-\frac{1}{2}}b_{-\vec{p},-\frac{1}{2}}^{\dag}-b_{\vec{p},\frac{1}{2}}^{\dag}b_{\vec{p},\frac{1}{2}}\Big).
\label{eq47}
\end{gather}

Now we derive the excited state wave function of superf\/luid Fermi gas on the noncommutative space.
The Fermionic quasi-particles must be created or annihilated in pairs, so the wave function of superf\/luid Fermi gas
with~$n$ pairs excited quasi-particles takes the form
\begin{gather}
|\psi^{\prime}_{n}\rangle=C\prod\limits_{n}d_{\vec{p}_{n},\sigma_{n}}^{\dag}d_{\vec{q}_{n},\sigma_{n}^{\prime}}^{\dag}|{\rm BCS}\rangle,
\label{eq48}
\end{gather}
where~$C$ is the normalization constant, and $d_{\vec{p}_{n},\sigma_{n}}^{\dag}$ is the creation operator of the~$n$th
pair of quasi-particles on the noncommutative space.

In general, a~single quasi-particle state created by $d_{\vec{p},{\sigma}}^{\dag}$ on the noncommutative space is
similar to the state created by $b_{\vec{p},{\sigma}}^{\dag}$ on the commutative space, but a~multi-particle state on
the noncommutative space has no corresponding observed quantities since it has a~dif\/ferent superposition rule from that
of the commutative case~\cite{APB2007,Basu}.
Fortunately, the excited state of multi Fermionic quasi-particles in the BCS-type superf\/luid (or superconductor) theory
is an exception.
In accordance with equation~\eqref{eq35}, we obtain
$d_{\vec{p}_{n},\sigma_{n}}^{\dag}d_{-\vec{p}_{n},\sigma_{n}^{\prime}}^{\dag}
=b_{\vec{p}_{n},\sigma_{n}}^{\dag}b_{-\vec{p}_{n},\sigma_{n}^{\prime}}^{\dag}$, which can be generalized to~$n$ pairs of
quasi-particles with~$n$ a~positive integer.
Then we simplify the wave function of superf\/luid Fermi gas with~$n$ pairs excited quasi-particles (see
equation~\eqref{eq48}) to be
\begin{gather}
|\psi_{n}\rangle=C\prod\limits_{n}b_{\vec{p}_{n},\sigma_{n}}^{\dag}b_{-\vec{p}_{n},\sigma_{n}^{\prime}}^{\dag}|{\rm
BCS}\rangle,
\label{eq49}
\end{gather}
which is similar to the wave function in the commutative case.
Note that $|{\rm BCS}\rangle$ is not only the ground state wave function of superf\/luid Fermi gas, but also the vacuum
state of quasi-particles, satisfying $b_{-\vec{p},-\frac{1}{2}}|{\rm BCS}\rangle=0$ and $b_{\vec{p},\frac{1}{2}}|{\rm
BCS}\rangle=0$.
Using equations~\eqref{eq46},~\eqref{eq47}, and~\eqref{eq49}, we then get the energy spectrum of superf\/luid Fermi gas{\samepage
\begin{gather}
E_{\theta} = \langle\psi_{n}| H_{\theta}^{\prime}| \psi_{n}\rangle
 = 2\sum\limits_{\vec{p}}\eta_{\vec{p}}\,v_{\vec{p}}^{2}+\sum\limits_{\vec{p}}\eta_{\vec{p}}\big(u_{\vec{p}}^{2}-v_{\vec{p}}^{2}\big)
\big(n_{\vec{p},\frac{1}{2}}+n_{\vec{p},-\frac{1}{2}}\big)
\nonumber
\\
\phantom{E_{\theta}=}
{} +\! \sum\limits_{\vec{p},\vec{p}^{\,\prime} }\tilde{U}(\vec{p}^{\,\prime} -\vec{p})
e^{-i\vec{p}^{\,\prime} \wedge\vec{p}}u_{\vec{p}}\,v_{\vec{p}}\,u_{\vec{p}^{\,\prime} }\,v_{\vec{p}^{\,\prime} }
\big(1-n_{\vec{p},\frac{1}{2}}-n_{-\vec{p},-\frac{1}{2}}\big)
\big(1-n_{\vec{p}^{\,\prime} ,\frac{1}{2}}-n_{-\vec{p}^{\,\prime} ,-\frac{1}{2}}\big),\!\!\!
\label{eq50}
\end{gather}
where $n_{\vec{p},{\sigma}}$ is the occupation number of quasi-particles.}

The occupation number of quasi-particles satisf\/ies the Fermi--Dirac distribution, and it does not depend on the
direction of momenta and spins.
Due to the isotropy of superf\/luid Fermi gas and the permutation symmetry of $\vec{p}$ and $\vec{p}^{\,\prime} $ in the
momentum summation, $\tilde{U}(\vec{p}^{\,\prime} -\vec{p})$ depends only on the absolute value of
$\vec{p}^{\,\prime} -\vec{p}$, not on the direction.
As a~result, the imaginary part of equation~\eqref{eq50} vanishes, equation~\eqref{eq50} is real in nature and can be
written as follows
\begin{gather}
E_{\theta} = 2\sum\limits_{\vec{p}}\eta_{\vec{p}}\,v_{\vec{p}}^{2}+\sum\limits_{\vec{p}}\eta_{\vec{p}}\big(u_{\vec{p}}^{2}-v_{\vec{p}}^{2}\big)
\big(n_{\vec{p},\frac{1}{2}}+n_{\vec{p},-\frac{1}{2}}\big)
\nonumber
\\
\phantom{E_{\theta}=}
{}+\sum\limits_{\vec{p},\vec{p}^{\,\prime} }\tilde{U}\cos\left(p_{i}\theta_{ij}p_{j}^{\prime}\right)u_{\vec{p}}\,v_{\vec{p}}\,u_{\vec{p}^{\,\prime} }\,
v_{\vec{p}^{\,\prime} }\big(1-n_{\vec{p},\frac{1}{2}}-
n_{\vec{p},-\frac{1}{2}}\big)\big(1-n_{\vec{p}^{\,\prime} ,\frac{1}{2}}-n_{\vec{p}^{\,\prime} ,-\frac{1}{2}}\big).
\label{eq51}
\end{gather}
This is the nonperturbative energy spectrum of superf\/luid Fermi gas on the noncommutative space.
Comparing it with the result on the commutative space, we f\/ind that each energy level has a~corrected factor of cosine
function of noncommutative parameters.
When $\theta_{ij}\rightarrow 0$, the energy spectrum reduces to the commutative case, which ensures the consistency of
our noncommutative generalization.

\subsection{Phase transition}

The energy gap is an important physical quantity that ref\/lects the superconductivity and superf\/luidity of systems, and
is intimately related to the critical temperature of phase transition.
Now we discuss the ef\/fect of space noncommutativity on the energy gap of superf\/luid Fermi gas and then determine how the
space noncommutativity modif\/ies the critical temperature of phase transition.
We shall f\/irst calculate the energy gap at zero temperature and then turn to the nonzero case in terms of the method of
quasi-particles~\cite{Lifshitz}.

Based on~\cite{Basu}, the weak attractive interaction in the superf\/luid Fermi gas can approximately be regarded as
a~trap, i.e., it equals a~negative constant $-U_0$ $(U_{0}>0)$ within a~small region near the Fermi surface but zero
beyond the small region.
Similar to the Debye frequency in solid state physics, there exists a~maximal excited frequency $\omega_{m}$ for
quasi-particles at a~f\/ixed temperature.
Assume that the attractive potential is nonzero within the energy shell $\hbar\omega_{m}$ around the Fermi
surface\footnote{For the sake of clearness, $\hbar$ is written explicitly.}, that is, when $|\eta_{\vec{p}}|\leq
\hbar\omega_{m}$ and $|\eta_{\vec{p}^{\,\prime}}|\leq \hbar\omega_{m}$, one has $\tilde{U}=-U_{0}$, otherwise
$\tilde{U}=0$.
Within this energy shell, considering that $|\vec {p}|$ is almost equal to $|\vec {p}_{\rm F}|$ and the three noncommutative
parameters ($\theta_{ij}$) are equal, denoted by~$\theta$, we can get $p_{i}\theta_{ij}p_{j}^{\prime}
\approx\kappa\theta$, where~$\kappa$ is proportional to the square of the Fermi momentum.
In consequence, the energy spectrum of the ground state (equation~\eqref{eq40}) can approximately be written as
\begin{gather*}
E_{\theta}^0=2\sum\limits_{\vec{p}}\eta_{\vec{p}}\,v_{\vec{p}}^{2}-U_0\cos{(\kappa\theta)}\sum\limits_{\vec{p},\vec{p}^{\,\prime} }
u_{\vec{p}}\,v_{\vec{p}}\,u_{\vec{p}^{\,\prime} }\,v_{\vec{p}^{\,\prime} }.
\end{gather*}

Because the BCS state is a~superf\/luid state at zero temperature, we adopt the common method~\cite{Lifshitz} to calculate
$u_{\vec{p}}$ and $v_{\vec{p}}$, i.e., to make the variation of the ground state ener\-gy~$E_{\theta}^0$ with respect to
$v_{\vec{p}}$ and $u_{\vec{p}}$, respectively, and then to require it be zero.
Considering the normalization condition $u_{\vec{p}}^{2}+v_{\vec{p}}^{2}=1$, we obtain
\begin{gather}
\frac{\partial E_{\theta}^0}{\partial v_{\vec{p}}}
=4\eta_{\vec{p}}\,v_{\vec{p}}-2U_0\cos{(\kappa\theta)}\left(\sum\limits_{\vec{p}^{\,\prime} }\,u_{\vec{p}^{\,\prime} }v_{\vec{p}^{\,\prime} }\right)
\frac{u_{\vec{p}}^{2}-v_{\vec{p}}^{2}}{\sqrt{1-v_{\vec{p}}^{2}}}=0.
\label{eq53}
\end{gather}
Using the def\/inition of energy gap at zero temperature
\begin{gather}
\Delta(0) \equiv U_0\cos{(\kappa\theta)}\sum\limits_{\vec{p}}u_{\vec{p}}\,v_{\vec{p}},
\label{eq54}
\end{gather}
we simplify equation~\eqref{eq53} as follows
\begin{gather}
2\eta_{\vec{p}}\,u_{\vec{p}}\,v_{\vec{p}}=\Delta(0)\big(u_{\vec{p}}^{2}-v_{\vec{p}}^{2}\big).
\label{eq55}
\end{gather}
Again using equation~\eqref{eq39}, we solve $u_{\vec{p}}$ and $v_{\vec{p}}$ from equation~\eqref{eq55}
\begin{gather}
u_{\vec{p}}^{2} = \frac{1}{2}\left(1+\frac{\eta_{\vec{p}}}{\sqrt{\Delta^{2}(0)+\eta_{\vec{p}}^{2}}}\right),
\label{eq56}
\\
v_{\vec{p}}^{2} = \frac{1}{2}\left(1-\frac{\eta_{\vec{p}}}{\sqrt{\Delta^{2}(0)+\eta_{\vec{p}}^{2}}}\right).
\label{eq57}
\end{gather}
Comparing equations~\eqref{eq56} and~\eqref{eq57} with the results on the commutative space~\cite{Lifshitz}, we see that
$u_{\vec{p}}^{2}$ and $v_{\vec{p}}^{2}$ take similar forms except for the appearance of the corrected factor of cosine
function in the def\/inition of the energy gap at zero temperature $\Delta(0)$ (see equation~\eqref{eq54}).

Substituting equations~\eqref{eq56} and~\eqref{eq57} into equation~\eqref{eq54}, we rewrite the energy gap at zero
temperature as
\begin{gather*}
\Delta(0)=\frac{1}{2}U_0\cos{(\kappa\theta)}\sum\limits_{\vec{p}}\frac{\Delta(0)}{\sqrt{\eta_{\vec{p}}^{2}+\Delta^{2}(0)}}.
\end{gather*}
Replacing the summation by integration in the above equation, we have
\begin{gather}
1=g(0)U_{0}\cos{(\kappa\theta)}\int_{0}^{\hbar\omega_{m}}\frac{d\eta_{\vec{p}}}{\sqrt{\eta_{\vec{p}}^{2}+\Delta^{2}(0)}},
\label{eq59}
\end{gather}
where $g(0)$ denotes the state density per unit energy interval around the Fermi surface.
Solving equation~\eqref{eq59}, we get the energy gap at zero temperature
\begin{gather}
\Delta(0)\approx2\hbar\omega_{m}\exp\left[{-\frac{1}{g(0)U_{0}\cos{(\kappa\theta)}}}\right].
\label{eq60}
\end{gather}
Compared with the result on the commutative space, an additional corrected factor of cosine function of noncommutative
parameters appears.
If the noncommutative parameters approach zero, the energy gap equation~\eqref{eq60} reduces to the result of the
commutative case.
Because the parameter $\kappa\theta$ is very small, the value of $\cos(\kappa\theta)$ is very close to but
less than one.
As a~result, the energy gap of superf\/luid Fermi gas on the noncommutative space is narrower than that of the commutative
case at zero temperature.

Now we discuss the energy gap at nonzero temperature, where the occupation number of quasi-particles
($n_{\vec{p},{\sigma}}$) is no longer vanishing.
Within the energy shell $\hbar\omega_{m}$ around the Fermi surface and under the approximation condition
$p_{i}\theta_{ij}p_{j}^{\prime} \approx\kappa\theta$, we simplify the energy spectrum of superf\/luid Fermi gas
(equation~\eqref{eq51}) to be
\begin{gather*}
E_{\theta} = 2\sum\limits_{\vec{p}}\eta_{\vec{p}}\,v_{\vec{p}}^{2}+\sum\limits_{\vec{p}}\eta_{\vec{p}}\big(u_{\vec{p}}^{2}-v_{\vec{p}}^{2}\big)
\big(n_{\vec{p},\frac{1}{2}}+n_{\vec{p},-\frac{1}{2}}\big)
\nonumber
\\
\phantom{E_{\theta}=}
{}-U_0\cos{(\kappa\theta)}\sum\limits_{\vec{p},\vec{p}^{\,\prime} }u_{\vec{p}}\,v_{\vec{p}}\,u_{\vec{p}^{\,\prime} }\,
v_{\vec{p}^{\,\prime} }\big(1-n_{\vec{p},\frac{1}{2}}-
n_{\vec{p},-\frac{1}{2}}\big)\big(1-n_{\vec{p}^{\,\prime} ,\frac{1}{2}}-n_{\vec{p}^{\,\prime} ,-\frac{1}{2}}\big).
\end{gather*}
Note that the expressions of $u_{\vec{p}}$ and $v_{\vec{p}}$ of excited states are similar to that of the ground state
(equations~\eqref{eq56} and~\eqref{eq57}) with only the replacement of $\Delta(0)$ by $\Delta(T)$, where the def\/inition
of the energy gap at nonzero temperature is
\begin{gather}
\Delta(T)\equiv U_0\cos{(\kappa\theta)}\sum\limits_{\vec{p}}u_{\vec{p}}\,v_{\vec{p}}
\big(1-n_{\vec{p},\frac{1}{2}}-n_{\vec{p},-\frac{1}{2}}\big).
\label{eq62}
\end{gather}

Next, we can calculate the critical temperature of phase transition and the energy gap at temperature~$T$ by following
the usual way~\cite{Lifshitz} because $\Delta(T)$ is same as that of the commutative case except for the coef\/f\/icient
$\cos{(\kappa\theta)}$.
In the following we just give the results but leave the details of derivations to the Appendix.

When the temperature rises, the superf\/luid state will transform into the ordinary state.
With the condition $\Delta(T_{\rm c})=0$, we can get the critical temperature
\begin{gather}
T_{\rm c}\approx 0.57\,\frac{\Delta(0)}{k_{\rm B}}.
\label{eq63}
\end{gather}
Since $\Delta(0)$ is smaller than the corresponding result of the commutative case, the critical temperature of phase
transition is lower in the noncommutative case.

If the temperature is extremely low ($T\ll T_{\rm c}$), the energy gap of superf\/luid Fermi gas on the noncommutative space
is
\begin{gather}
\Delta(T)\approx\Delta(0)\left[1-\sqrt{\frac{2\pi k_{\rm B}T}{\Delta(0)}}\exp\left(-\frac{\Delta(0)}{k_{\rm B}T}\right)\right],
\qquad
T\ll T_{\rm c}.
\label{eq64}
\end{gather}

In the vicinity of the phase transition temperature, i.e.,~$T$ is close to but less than $T_{\rm c}$, the energy gap of
superf\/luid Fermi gas takes the form,
\begin{gather}
\Delta(T)\approx 1.74\, \Delta(0)\left(1-\frac{T}{T_{\rm c}}\right)^{\frac{1}{2}},
\qquad
T_{\rm c}-T\ll T_{\rm c}.
\label{eq65}
\end{gather}
If the temperature is higher than the critical temperature $T_{\rm c}$, the energy gap disappears and the superf\/luid Fermi
gas changes into the ordinary state.
We can see from equations~\eqref{eq64} and~\eqref{eq65} that the energy gap on the noncommutative space is narrower than
that on the commutative space if $T<T_{\rm c}$, which is in agreement with the result of superconductor on the Moyal
plane~\cite{Basu}.
This result is understandable.
Because the interaction in the superf\/luid Fermi gas of atoms with pairs of quasi-particles of opposite momenta and spins
is of BCS-type, which coincides with the interaction in the superconductor~\cite{Basu}, the space noncommutativity leads
to a~reduction of the energy gap in both works.
However, the dif\/ference between them is also obvious.
Our result is nonperturbative and can describe both the ground state and excited states of the superf\/luid Fermi gas of
atoms, while~\cite{Basu} gives a~perturbative correction only for the ground state of the superconductor.

\section{Conclusion and perspective}
\label{Section4}

Considering the special interaction in the BCS-type superf\/luid Fermi gas on the canonical noncommutative space, we get
the nonperturbative energy spectrum of the system.
Compared with the result on the commutative space, each energy level contains an additional corrected factor of cosine
function of noncommutative parameters.
When the temperature rises, the phase transition from the superf\/luid to ordinary f\/luid states occurs on the
noncommutative space, but the critical temperature of phase transition is lower than that of the commutative case.
Our result may be regarded as a~theoretical prediction for the existence of space noncommutativity in future
experiments.

We would like to make some comments on the Moyal star product.
It was shown~\cite{PBasu} that the Moyal star product is somewhat unphysical, in contrast to the Voros star product, as
the former does not conform to a~positive-operator-valued measure (POVM).
Consequently, it was suggested~\cite{FGScholtz} that one can compute Connes distance function in the Voros basis and not
in the Moyal basis.
Further development~\cite{YCDevi} showed that the Voros basis can also be constructed in the odd 3-dimensional space,
where one introduces the so-called ``quasi-commutative" bases and can recover the usual Bose/Fermi symmetry and the
undeformed (canonical) commutation relations in place of equations~\eqref{eq17}--\eqref{eq19} of the present paper.
Not only that, the relations in equation~\eqref{eq28} can be seen as the connection between the creation (annihilation)
operators of Fermionic quasi-particles in the twisted and the quasi-commutative bases.
Therefore, it is an interesting issue to investigate the energy spactrum and phase transition of the superf\/luid Fermi
gas of atoms in terms of the Voros star product, which will be done in our future work under the particular
consideration of the characteristic of the Voros basis~-- the non-orthogonality.

Moreover, there are three noncommutative algebras that are commonly used in physics, that is, the canonical,
Lie-algebraic, and quadratic forms.
In this paper we focus only on the superf\/luid Fermi gas on the canonical noncommutative space, we shall try to
investigate superconductivity and superf\/luidity on a~kind of Lie-algebraic noncommutative spacetimes, such as the
important~$\kappa$-deformed Minkowski spacetime~\cite{Ame,Ame+}.
This specif\/ic Lie-algebraic noncommutative spacetime has attracted a~lot of attentions since it is intimately related to
the special relativity with two constants.
The key point of such an investigation is to work out the anti-commutation relations similar to
equations~\eqref{eq17},~\eqref{eq18}, and~\eqref{eq19} for the~$\kappa$-deformed Minkowski spacetime in light of the
twisted deformed Hopf algebra~\cite{Bu}, which is now under consideration.

\appendix

\section{Derivations of equations~(\ref{eq63}), (\ref{eq64}) and (\ref{eq65})}

Now we rewrite the energy gap at temperature~$T$ (equation~\eqref{eq62}) within the energy shell $|\eta_{\vec{p}}|\leq
\hbar\omega_{m}$ above the Fermi surface.
Substituting $u_{\vec{p}}$ and $v_{\vec{p}}$ (see equations~\eqref{eq56} and~\eqref{eq57}) into equation~\eqref{eq62},
we obtain
\begin{gather}
\frac{1}{2}U_0\cos{(\kappa\theta)}\sum\limits_{\vec{p}} \frac{1-n_{\vec{p},\frac{1}{2}}-n_{\vec{p},-\frac{1}{2}}}
{\sqrt{\Delta^{2}(T)+\eta_{\vec{p}}^{2}}}=1.
\label{eq66}
\end{gather}
In a~thermal equilibrium state, the occupation number of quasi-particles does not depend on spins and can be described
by the Fermi--Dirac distribution (with a~vanishing chemical potential) as follows
\begin{gather*}
n_{\vec{p}} \equiv n_{\vec{p},\frac{1}{2}}= n_{\vec{p},-\frac{1}{2}}=
\frac{1}{\exp\left(\frac{\varepsilon}{k_{\rm B}T}\right)+1},
\end{gather*}
where~$\varepsilon$ is the energy of a~single quasi-particle, $k_{\rm B}$ is the Boltzmann constant, and~$T$ is tem\-pe\-ra\-ture.
When the summation is replaced by integration, equation~\eqref{eq66} becomes
\begin{gather}
\frac{1}{2}U_{0}\cos(\kappa\theta)\int \frac{1-2n_{\vec{p}}}
{\sqrt{\Delta^{2}(T)+\eta_{\vec{p}}^{2}}}\frac{d^{3}p}{(2\pi\hbar)^{3}}=1.
\label{eq68}
\end{gather}
Since the integration region is very small, i.e., $|\eta_{\vec{p}}|\leq \hbar\omega_{m}$, the variable of integration
can be changed to be $\eta_{\vec{p}}$ in terms of equation~\eqref{eq39}.
Through def\/ining an equivalent quantum state density $g(0)$ around the Fermi surface ($|\eta_{\vec{p}}|\leq
\hbar\omega_{m}$) which is used to approximately represent the state density of the attractive region, we thus rewrite
equation~\eqref{eq68} to be
\begin{gather*}
\frac{1}{2}g(0)U_{0}\cos(\kappa\theta)\int_{-\hbar\omega_{m}}^{\hbar\omega_{m}}
\frac{1-2n_{\vec{p}}}{\sqrt{\Delta^{2}(T)+\eta_{\vec{p}}^{2}}} d\eta_{\vec{p}}=1.
\end{gather*}
For the sake of convenience in the following calculation, we again simplify the above equation to be{\samepage
\begin{gather}
g(0)U_{0}\cos(\kappa\theta)\int_{0}^{\hbar\omega_{m}} \frac{1-2n(\varepsilon)}{\sqrt{\Delta^{2}(T)+\eta_{\vec{p}}^{2}}}
d\eta_{\vec{p}}=1,
\label{eq70}
\end{gather}
where $n(\varepsilon) \equiv n_{\vec{p}}=\frac{1}{\exp\big(\frac{\varepsilon}{k_{\rm B}T}\big)+1}$
and $\varepsilon=\sqrt{\Delta^{2}(T)+\eta_{\vec{p}}^{2}}$.}

\subsection{Derivation of equation~(\ref{eq63})}

We calculate the critical temperature of phase transition using the condition $\Delta(T_{\rm c})=0$ under which
equation~\eqref{eq70} becomes
\begin{gather}
1=g(0)U_{0}\cos{(\kappa\theta)}\int_{0}^{\hbar\omega_{m}}d\eta_{\vec{p}}\,
\frac{\tanh\left(\frac{1}{2}\beta_{\rm c}\eta_{\vec{p}}\right)}{\eta_{\vec{p}}} =
g(0)U_{0}\cos{(\kappa\theta)}\int_{0}^{\frac{1}{2}\beta_{\rm c}\hbar\omega_{m}}dx\, \frac{\tanh x}{x},
\label{eq71}
\end{gather}
where $\beta_{\rm c}\equiv (k_{\rm B}T_{\rm c})^{-1}$ and $x\equiv \frac{1}{2}\beta_{\rm c}\eta_{\vec{p}}$.
Doing the integration by parts, we can get
\begin{gather*}
1=g(0)U_{0}\cos{(\kappa\theta)}\left(\ln x\tanh x \Bigg|_{0}^{\frac{1}{2}\beta_{\rm c}\hbar\omega_{m}}
-\int_{0}^{\frac{1}{2}\beta_{\rm c}\hbar\omega_{m}}dx\frac{\ln x}{\cosh^{2}x}\right).
\end{gather*}
Since the integral converges quickly, and $\hbar\omega_{m}\gg k_{\rm B}T_{\rm c}$, the upper limit of integral can be changed to
be inf\/inity, then the above equation becomes
\begin{gather}
1=g(0)U_{0}\cos{(\kappa\theta)}\left[\ln \left(\frac{\hbar\omega_{m}}{2k_{\rm B}T_{\rm c}}\right) -\int_{0}^{\infty}dx\frac{\ln
x}{\cosh^{2}x}\right].
\label{eq73}
\end{gather}
Using the Euler integration formula
\begin{gather*}
\int_{0}^{\infty}dx\frac{\ln x}{\cosh^{2}x}=-\ln\frac{4e^{\gamma}}{\pi},
\end{gather*}
where $\gamma \approx 0.5772$ is the Euler constant, we simplify equation~\eqref{eq73} to be
\begin{gather}
k_{\rm B}T_{\rm c}=\frac{e^{\gamma}}{\pi}\,2\hbar\omega_{m}\exp\left[-\frac{1}{g(0)U_{0}\cos{(\kappa\theta)}}\right].
\label{eq75}
\end{gather}
Taking into account equation~\eqref{eq60}, we at last reach equation~\eqref{eq63}
\begin{gather}
T_{\rm c}=\frac{e^{\gamma}}{\pi}\frac{\Delta(0)}{k_{\rm B}}\approx 0.57\,\frac{\Delta(0)}{k_{\rm B}}.
\label{eq76}
\end{gather}

\subsection{Derivation of equation~(\ref{eq64})}

First change equation~\eqref{eq70} into the following form,
\begin{gather}
-1+g(0)U_{0}\cos{(\kappa\theta)}\int_{0}^{\hbar\omega_{m}}\frac{d\eta_{\vec{p}}}{{\sqrt{\Delta^{2}(T)
+\eta_{\vec{p}}^{2}}}}=2g(0)U_{0}\cos{(\kappa\theta)}\int_{0}^{\hbar\omega_{m}}
\frac{n(\varepsilon)d\eta_{\vec{p}}}{\sqrt{\Delta^{2}(T) +\eta_{\vec{p}}^{2}}}.
\label{eq77}
\end{gather}
Due to $\hbar\omega_{m}\gg \Delta(0)$, the second term on the left side of equation~\eqref{eq77} can be written as
\begin{gather}
\int_{0}^{\hbar\omega_{m}} \frac{d\eta_{\vec{p}}}{{\sqrt{\Delta^{2}(T)+\eta_{\vec{p}}^{2}}}} =
\ln\frac{2\hbar\omega_{m}}{\Delta(T)}.
\label{eq78}
\end{gather}
Note that equation~\eqref{eq60} implies
\begin{gather}
1=g(0)U_{0}\cos{(\kappa\theta)}\ln\frac{2\hbar\omega_{m}}{\Delta(0)}.
\label{eq79}
\end{gather}
Substituting equations~\eqref{eq78} and~\eqref{eq79} into the left side of equation~\eqref{eq77}, we see that the left
side equals $g(0)U_{0}\cos{(\kappa\theta)}\ln\frac{\Delta(0)}{\Delta(T)}$.
Again substituting $n(\varepsilon)$ (see the notations under equation~\eqref{eq70}) into the right side of
equation~\eqref{eq77}, we can rewrite equation~\eqref{eq77} to be
\begin{gather}
\ln\frac{\Delta(0)}{\Delta(T)}=2\int_{0}^{\hbar\omega_{m}}d\eta_{\vec{p}}\,\frac{1} {\sqrt{\eta_{\vec{p}}^{2}
+\Delta^{2}(T)}}\frac{1}{\exp\left(\beta\sqrt{\eta_{\vec{p}}^{2} +\Delta^{2}(T)}\right)+1},
\label{eq80}
\end{gather}
where $\beta\equiv(k_{\rm B}T)^{-1}$.
At an extremely low temperature ($T\ll T_{\rm c}$), $\Delta(T)$ on the right side of equation~\eqref{eq80} can be replaced
by $\Delta(0)$, the above equation then becomes
\begin{gather*}
\ln\frac{\Delta(T)}{\Delta(0)}=-2\int_{0}^{\hbar\omega_{m}}d\eta_{\vec{p}}\,
\frac{\exp\left(-\beta\sqrt{\eta_{\vec{p}}^{2} +\Delta^{2}(0)}\right)}{\sqrt{\eta_{\vec{p}}^{2} +\Delta^{2}(0)}}.
\end{gather*}
Considering the condition $\beta\Delta(0)\gg 1$, we see that the exponential function decays quickly, so the energy
region around $\eta_{\vec{p}}\approx 0$ makes the main contribution to the integration.
In addition, $\sqrt{\eta_{\vec{p}}^{2}+\Delta^{2}(0)}$ can be expanded in $\frac{\eta_{\vec{p}}}{\Delta(0)}$ as a~small
quantity.
Taking into account $\hbar\omega_{m}\gg k_{\rm B}T$, we make an approximation that the integral upper limit can be replaced
by inf\/inity.
With all these considerations, we simplify the above equation to be
\begin{gather}
\ln\frac{\Delta(T)}{\Delta(0)} = -\frac{2}{\Delta(0)}\int_{0}^{\infty}d\eta_{\vec{p}}\,
\exp\left[-\frac{\Delta(0)}{k_{\rm B}T}\left(1+\frac{\eta_{\vec{p}}^{2}}{2\Delta^{2}(0)}\right)\right]
\nonumber
\\
\phantom{\ln\frac{\Delta(T)}{\Delta(0)}}
 = -\sqrt{\frac{2\pi k_{\rm B}T}{\Delta(0)}}\exp\left(-\frac{\Delta(0)}{k_{\rm B}T}\right).
\label{eq82}
\end{gather}
If $T\ll T_{\rm c}$, we have $\Delta(0)-\Delta(T)\ll \Delta(0)$ and therefore simplify the left side of
equation~\eqref{eq82} to be
\begin{gather}
\ln\frac{\Delta(T)}{\Delta(0)}=\ln\left(1+\frac{\Delta(T)-\Delta(0)}{\Delta(0)}\right)
\approx\frac{\Delta(T)-\Delta(0)}{\Delta(0)}.
\label{eq83}
\end{gather}
Combining equation~\eqref{eq82} with equation~\eqref{eq83}, we f\/inally reach equation~\eqref{eq64}
\begin{gather*}
\Delta(T)\approx\Delta(0)\left[1-\sqrt{\frac{2\pi k_{\rm B}T}{\Delta(0)}}\exp\left(-\frac{\Delta(0)}{k_{\rm B}T}\right)\right].
\end{gather*}

\subsection{Derivation of equation~(\ref{eq65})}

Due to $n(\varepsilon)+n(-\varepsilon)=1$, equation~\eqref{eq70} can be written as
\begin{gather}
g(0)U_{0}\cos{(\kappa\theta)}\int_{0}^{\hbar\omega_{m}}d\eta_{\vec{p}}\,
\frac{n(-\varepsilon)-n(\varepsilon)}{\varepsilon}=1.
\label{eq85}
\end{gather}
Using the Poisson summation formula
\begin{gather*}
n(\varepsilon)=k_{\rm B}T\lim_{\lambda\rightarrow
0^{+}}\sum\limits_{k=-\infty}^{\infty}\frac{e^{i\omega_{k}\lambda}}{i\omega_{k}-\varepsilon},
\end{gather*}
where $\omega_{k}=(2k+1)\pi k_{\rm B}T$, $k$ is an integer, we have
\begin{gather}
\frac{n(-\varepsilon)-n(\varepsilon)}{\varepsilon}=4k_{\rm B}T\sum\limits_{k=0}^{\infty}
\frac{1}{\omega_{k}^{2}+\eta_{\vec{p}}^{2}+\Delta^{2}(T)}.
\label{eq87}
\end{gather}
Substituting the above equation into equation~\eqref{eq85}, we get
\begin{gather}
4k_{\rm B}T g(0)U_{0}\cos{(\kappa\theta)}\int_{0}^{\hbar\omega_{m}}d\eta_{\vec{p}}
\sum\limits_{k=0}^{\infty}\frac{1}{\omega_{k}^{2}+\eta_{\vec{p}}^{2}+\Delta^{2}(T)}=1.
\label{eq88}
\end{gather}
Because the energy gap ${\Delta}(T)$ is very narrow when~$T$ is around the critical temperature, it can be regarded as
a~small quantity, so the following series expansion is given
\begin{gather*}
\frac{1}{\omega_{k}^{2}+\eta_{\vec{p}}^{2}+\Delta^{2}(T)}=\frac{1}{\big(\omega_{k}^{2}+\eta_{\vec{p}}^{2}\big)}
-\frac{\Delta^{2}(T)}{\big(\omega_{k}^{2}+\eta_{\vec{p}}^{2}\big)^{2}}+\cdots.
\end{gather*}
Taking only the f\/irst two terms, and substituting them into equation~\eqref{eq88}, we have
\begin{gather}
4k_{\rm B}T g(0)U_{0}\cos{(\kappa\theta)}\left[\sum\limits_{k=0}^{\infty}\int_{0}^{\hbar\omega_{m}}
\frac{d\eta_{\vec{p}}}{\omega_{k}^{2}+\eta_{\vec{p}}^{2}} -\Delta^{2}(T)
\sum\limits_{k=0}^{\infty}\int_{0}^{\hbar\omega_{m}}
\frac{d\eta_{\vec{p}}}{\big(\omega_{k}^{2}+\eta_{\vec{p}}^{2}\big)^{2}} \right]=1.
\label{eq90}
\end{gather}
Because the energy gap ${\Delta}(T)$ is a~small quantity, equation~\eqref{eq87} can approximately be written~as
\begin{gather}
4k_{\rm B}T \sum\limits_{k=0}^{\infty}
\frac{1}{\omega_{k}^{2}+\eta_{\vec{p}}^{2}}=\frac{\tanh(\frac{1}{2}\beta\eta_{\vec{p}})}{\eta_{\vec{p}}}.
\label{eq91}
\end{gather}
Substituting equation~\eqref{eq91} into equation~\eqref{eq90} and using equations~\eqref{eq71}--\eqref{eq75}, we simplify
equation~\eqref{eq90} to be
\begin{gather}
\ln\left(\frac{T}{T_{\rm c}}\right) = -4k_{\rm B}T\Delta^{2}(T)\sum\limits_{k=0}^{\infty}
\int_{0}^{\infty}d\eta_{\vec{p}}\,\frac{1}{\left(\omega_{k}^{2}+\eta_{\vec{p}}^{2}\right)^{2}}
 = -\frac{7}{8}\zeta(3)\left({\frac{\Delta(T)}{\pi k_{\rm B}T}}\right)^{2},
\label{eq92}
\end{gather}
where the Riemann zeta function takes the form
\begin{gather*}
\zeta(3)=\frac{8}{7}\sum\limits_{k=0}^{\infty}\frac{1}{(2k+1)^{3}}.
\end{gather*}
When~$T$ is close to but less than the critical temperature $T_{\rm c}$, i.e., $T_{\rm c}-T \ll T_{\rm c}$, the left side of
equation~\eqref{eq92} can approximately be written as
\begin{gather}
\ln\left(\frac{T}{T_{\rm c}}\right)=\ln\left(1-\frac{T_{\rm c}-T}{T_{\rm c}}\right)\approx -\left(1-\frac{T}{T_{\rm c}}\right).
\label{eq94}
\end{gather}
Substituting equations~\eqref{eq92} and~\eqref{eq94} into~\eqref{eq76}, we f\/inally reach~\eqref{eq65}
\begin{gather*}
\Delta(T)\approx\sqrt{\frac{8}{7\xi(3)}}e^{\gamma}\Delta(0)\left(1-\frac{T}{T_{\rm c}}\right)^{\frac{1}{2}}
\approx1.74\,\Delta(0)\left(1-\frac{T}{T_{\rm c}}\right)^{\frac{1}{2}},
\qquad
T_{\rm c}-T\ll T_{\rm c}.
\end{gather*}

\subsection*{Acknowledgments}
Y.-G.~Miao would like to thank J.-X.~Lu of the Interdisciplinary Center for Theoretical Study (ICTS), University of
Science and Technology of China (USTC) for warm hospitality where part of the work was performed.
This work was supported in part by the National Natural Science Foundation of China under grant No.~11175090 and by the
Ministry of Education of China under grant No.~20120031110027.
At last, the authors would like to thank the anonymous referees and the editor for their helpful comments that indeed
improve this work greatly.

\pdfbookmark[1]{References}{ref}
\LastPageEnding

\end{document}